\documentclass[9pt,twocolumn,twoside]{opticajnl}
\journal{opticajournal} 

\setboolean{shortarticle}{true}

\usepackage{lineno}
\usepackage{upgreek}

\title{High Contrast Nulling in Photonic Meshes Through Architectural Redundancy}

\author[1,*]{Carson G. Valdez}
\author[1]{Zhanghao Sun}
\author[1]{Anne R. Kroo}
\author[1]{David A. B. Miller}
\author[1]{Olav Solgaard}

\affil[1]{Electrical Engineering, Stanford University, 348 Via Pueblo, Stanford, CA 94305}

\affil[*]{carsongv@stanford.edu}

\begin{abstract}

We demonstrate a silicon photonic architecture comprised of Double Mach-Zehnder Interferometers (DMZIs) designed for high-contrast photonic applications. This configuration significantly enhances the achievable extinction ratio of photonic integrated circuits (PICs), reaching levels exceeding $80\ dB$. By leveraging the tunable properties of DMZIs and implementing a systematic configuration algorithm, the proposed mesh effectively compensates for fabrication imperfections and mitigates non-idealities such as back reflections. Experimental validation on a silicon-on-insulator platform demonstrates the potential of this approach for applications requiring high contrast nulling such as astronomical sensing.

\end{abstract}

\setboolean{displaycopyright}{false} 

\begin{document}

\maketitle


\textbf{Introduction.} Interferometric sensing is a cornerstone technique across optics and photonics owing largely to the high sensitivities that can be achieved\cite{Wen2022, Huang2021}. However, large interferometric systems are often vulnerable to instability resulting from thermal fluctuations and vibrations, among other noise sources \cite{Cho2009, Hati2008}. In bulk fiber optic systems, it is often difficult to stabilize and control more than a few interferometers at once. Integrated photonics has offered a pathway toward systems containing dozens to hundreds of interferometers in a compact footprint often much less than $1\ cm^2$ in area \cite{Bogaerts2020, Capmany2017}. 

Photonic integrated circuits fabricated in silicon, silicon nitride, or similar integrated platforms improve the scalability of large interferometric networks by simultaneously reducing the footprint of individual devices and providing robustness to environmental changes\cite{Bogaerts2018}. PICs are generally robust to dynamic strain-induced phase changes that may arise from external vibrations\cite{Humphreys2014}. Additionally, their compact form factor allows for efficient temperature stabilization via thermo-electric cooling (TEC)\cite{Enright2014}. These benefits directly address the challenges that prevent the scaling of fiber-based and free space-based interferometers where the optical paths may have to be distributed over lengths up to many meters.

Integrated photonic meshes - networks of interconnected Mach-Zehnder Interferometers (MZIs) - have previously been employed for complex wavefront sensing\cite{Sun2023}, optical computation \cite{Pai2023,Cheng2021}, and optical domain signal processing \cite{Annoni2017,Milanizadeh2022}. Recently, photonic meshes have been proposed to aid in the detection and characterization of Earth-like exoplanets. However, there remains a performance gap to be bridged for high-contrast applications such as astronomical interferometry and coronagraphy\cite{Vaughan2023,Sirbu2024}. 

A key metric for the performance of these meshes is the extinction ratio between channels. In a typical integrated photonic MZI the extinction ratio is limited by beam splitter errors caused by fabrication imperfections\cite{Pai2019}. A technique proposed to compensate for this fabrication error is to use individual MZI's as tunable beam-splitting components for a larger interferometer. This Double Mach-Zehnder Interferometer (DMZI) technique has demonstrated that the addition of these programmable elements provides additional robustness against errors\cite{Miller2015, Wilkes2016, Wang20}. 

We have expanded upon prior work by constructing a three-by-three triangular mesh of DMZIs. Here, successive elements of a diagonal line within the mesh are configured to further filter leakage power from previous elements. Additional elements not along the diagonal line are used to compensate for non-idealities, such as back reflections from coupling interfaces. To operate the mesh and determine an optimal state, we employ self-configuration algorithms that have previously been developed to operate photonic meshes\cite{Miller2015}. Leveraging the additional degrees of freedom of our circuit architecture, we achieve a single channel extinction ratio in excess of $80\ dB$.

\begin{figure}[ht]
\centering
{\includegraphics[width=\linewidth]{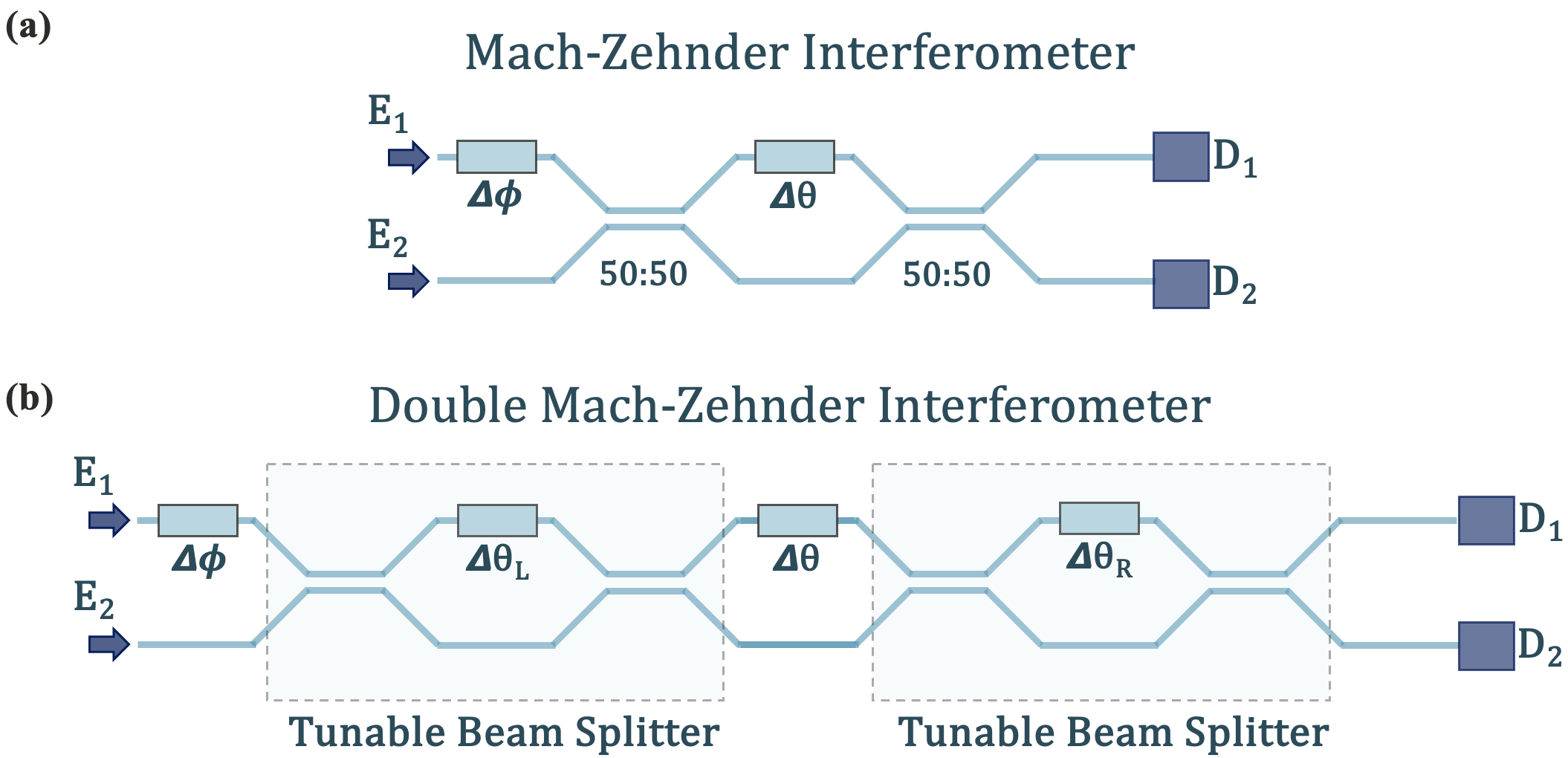}}
\caption{\textbf{(a)}. Schematic diagram of an ideal Mach-Zehnder interferometer with perfect split ratios. Phase shifters $\Delta{\theta}$ and $\Delta{\phi}$ set the matrix elements of the $2x2$ transformation applied to input fields $E_1$ and $E_2$. The output fields are measured at detectors $D_1$ and $D_2$. \textbf{(b)}. Schematic diagram of a Double Mach-Zehnder Interferometer with tunable beam splitters. Phase shifters $\Delta{\theta}_L$ and $\Delta{\theta}_R$ control the split ratio of the left and right beam splitter respectively.}
\label{Schematic}
\end{figure}

\textbf{Methods.} A typical Mach-Zehnder interferometer as shown in the schematic diagram in Figure \ref{Schematic}\color{blue}(a) \color{black} is comprised of an external phase shifter ($\Delta{\phi}$), a nominal $50:50$ beam splitter, an internal phase shifter ($\Delta{\theta}$), and an additional nominal $50:50$ beam splitter. Adjustment of the external phase shifter ($\Delta{\phi}$) controls the relative phase between fields at each of the input ports. Adjustment of the internal phase shifter ($\Delta{\theta}$) controls the degree of interference within the interferometer. Together the two degrees of freedom enable the application of an arbitrary, linear, unitary transformation of the complex input fields \cite{Miller2013}.

The corresponding transfer matrix assuming ideal components is given by equation \ref{eq:refname1} where the B matrices describe ideal beam splitters and the T matrices describe the effect of the phase shifters.

\begin{multline}
T(\theta,\phi) = BT_{\theta}BT_{\phi} \\
 = \frac{1}{\sqrt{2}}\begin{bmatrix} 1 & j \\ j & 1 \end{bmatrix}
\begin{bmatrix} e^{j\Delta\theta} & 0 \\ 0 & 1 \end{bmatrix}
\frac{1}{\sqrt{2}}\begin{bmatrix} 1 & j \\ j & 1 \end{bmatrix}
\begin{bmatrix} e^{j\Delta\phi} & 0 \\ 0 & 1 \end{bmatrix} \\
 = je^{j\frac{\Delta\theta}{2}}\begin{bmatrix} e^{j\Delta\phi}sin(\frac{\Delta\theta}{2}) & cos(\frac{\Delta\theta}{2}) \\  e^{j\Delta\phi}cos(\frac{\theta}{2}) & -sin(\frac{\Delta\theta}{2}) \end{bmatrix}
\label{eq:refname1}
\end{multline}

However, in practical implementations each of the beam splitter suffers some split ratio error ($\sigma$) introduced by small imperfections in the fabrication process. The corresponding transfer matrix for a practical beam splitter is given by equation \ref{eq:refname2}. 

\setlength{\arraycolsep}{5.0pt} 
\medmuskip = 0.0mu 

\begin{equation}
B(\sigma) =
\frac{1}{\sqrt{2}}\begin{bmatrix} \sqrt{1 + \sigma} & j\sqrt{1 - \sigma} \\  j\sqrt{1 - \sigma} & \sqrt{1 + \sigma} \end{bmatrix}
\label{eq:refname2}
\end{equation}




The subset of linear transformations that can be achieved by a practical MZI is limited by the severity of the split ratio error. In particular this limits the achievable extinction ration between the outputs of the MZI. As a direct consequence, when each of the phase shifters are set to direct all optical power to a single output port, there is a non-zero leakage to the undesired output port \cite{Pai2019}.

The Double Mach-Zehnder Interferometer reduces this leakage power by adding device complexity in order to compensate for fabrication imperfections. Figure \ref{Schematic}\color{blue}(b) \color{black} gives the schematic diagram of a DMZI comprised of an external phase shifter ($\Delta{\phi}$), an MZI acting as a tunable beam splitter, an internal phase shifter ($\Delta{\theta}$), and an additional MZI acting as a tunable beam splitter. Each of the individual MZIs is itself controlled by an internal phase shifter ($\Delta{\theta_L}$ and $\Delta{\theta_R}$). By adjusting $\Delta{\theta_L}$ and $\Delta{\theta_R}$ the split ratio of the tunable beam splitters can be driven towards $50:50$, effectively deepening the achievable rejection. 

Algorithms have been developed to configure the phase settings applied to $\Delta{\theta_L}$ and $\Delta{\theta_R}$ to achieve an ideal $50:50$ splitter \cite{Miller2015}. With a constant input power to a single input of the DMZI, the optimal split ratio settings of the individual MZI components are determined according to the following method:
\begin{enumerate}
\item Adjust $\Delta\theta$ to minimize the power at detector 2.
\item Adjust $\Delta{\theta_L}$ and $\Delta{\theta_R}$ with the same polarity ($+,+$) or ($-,-$) to further minimize the power at detector 2.
\item Adjust $\Delta\theta$ to minimize the power at detector 1.
\item Adjust $\Delta{\theta_L}$ and $\Delta{\theta_R}$ with opposite polarities ($+,-$) or ($-,+$) to further minimize the power at detector 1.
\item Iterate steps 1-4 until the phase shifter settings for $\Delta{\theta_L}$ and $\Delta{\theta_R}$ converge.
\end{enumerate}

After convergence this algorithm will have set the tunable beam splitter to the optimal state that can be achieved given the constraints of the system. This technique has been demonstrated to achieve extinction ratios as high as $60.5\ dB$ \cite{Wilkes2016}, representing several orders of magnitude improvement over traditional MZIs. While these improvements are substantial, the leakage power must be reduced further for applications requiring high-contrast nulling such as photonic coronagraphy. It has been suggested that direct detection of 'Earth-like' exo-planets would require contrasts on the order of $10^{-10}$  \cite{Vaughan2023, Sirbu2024}.

We note that other interesting approaches for improving performance of MZIs in interferometer meshes have been proposed that are simpler than the full DMZI discussed here, involving fewer additional beam splitters and phase shifters\cite{Hamerly22, Suzuki15, Wang20}. However, these do not support the simple convergent optimization algorithm discussed above, though they may support other general optimization algorithms. 

\begin{figure}[ht]
\centering
{\includegraphics[width=\linewidth]{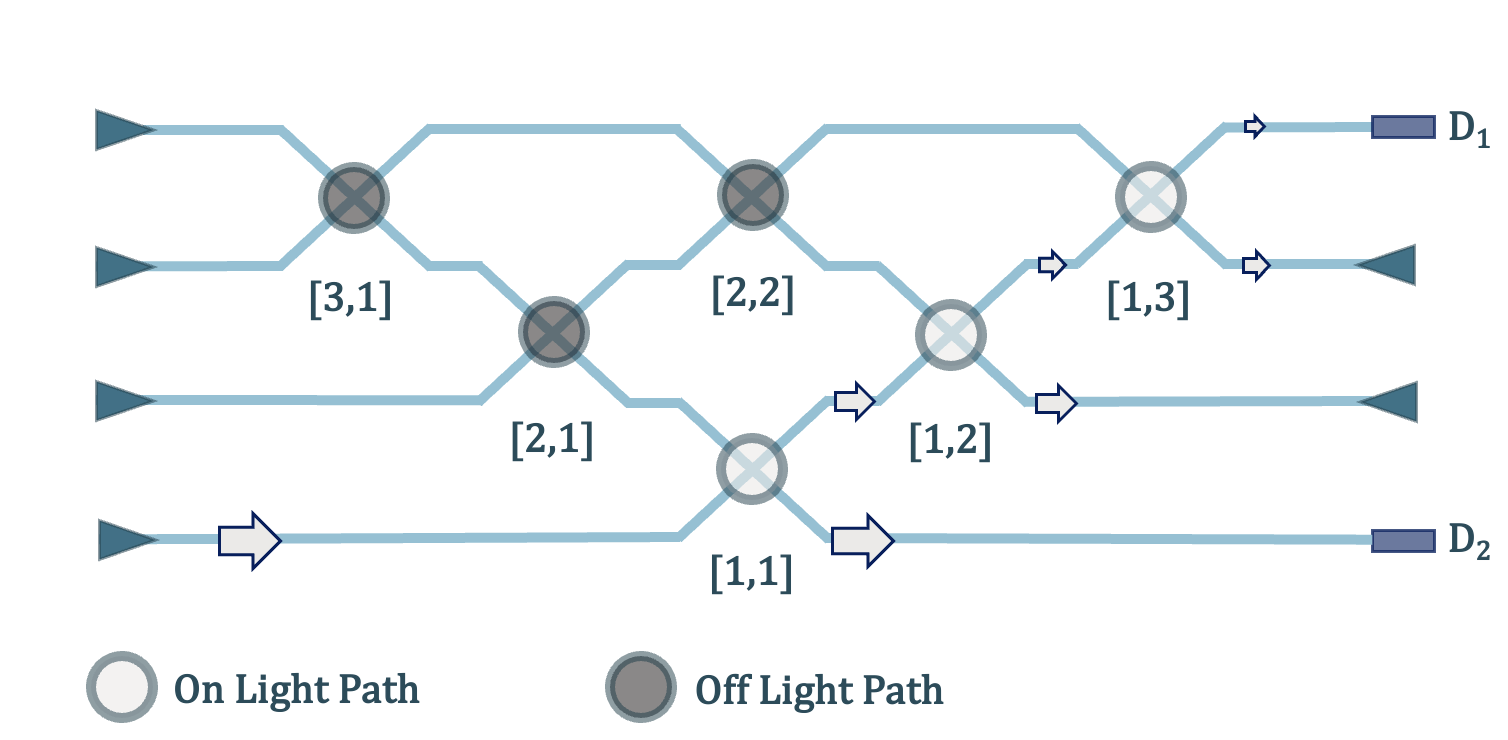}}
\caption{Schematic diagram of a 3x3 triangular mesh where each node is a Double Mach-Zehnder Interferometer. Subsequent elements of a single diagonal line along the light path are used to successively filter leakage from the previous elements. Additionally elements not along the light path are configured to reduce the impact of back reflections.}
\label{Mesh}
\end{figure}

To compensate for the remaining leakage power through a single DMZI, we have constructed a three-by-three triangular mesh of DMZIs. A schematic diagram of the proposed architecture is given in figure \ref{Mesh}. The triangular mesh is characterize by consecutive diagonal lines of DMZIs and allows for the generation of an arbitrary four-by-four linear, unitary transfer matrix.

Here we use a three-element diagonal line, containing the DMZIs marked in figure \ref{Mesh} as being on the light path (DMZIs $[1,1]$, $[1,2]$, and $[1,3]$), to successively reject power directed to detector $D_1$. In the ideal case the remaining devices marked as being off of the light path should have no impact on power measured at detector $D_1$. In practice however, we observe that optimization of these off light path devices does improve rejection at the output. We attribute these effects to non-negligible reflections from the coupling interfaces, allowing light to propagate along unexpected paths and interact with devices $[2,1]$, $[2,2]$, and $[3,1]$.


Each DMZI within the mesh is configured according to the algorithm detailed above. In order to configure each DMZI independently of the other devices a particular order must be followed \cite{Miller2015}. Optical power is initially input into the bottom port of the mesh allowing device $[1,1]$ to be configured. Using device $[1,1]$, the power is then directed to the input of device $[1,2]$ for configuration and so on to device $[1,3]$. Once this diagonal line of elements has completed configuration, each element is set to act as a transparent layer, allowing light to propagate through without altering its phase or intensity. The remaining devices then form a smaller 2x2 triangular mesh which is configured according to the same principle. This process is repeated until each element has been configured.


\textbf{Results and Discussion.} Our photonic integrated circuits have been fabricated on $220\ nm$ thick silicon-on-insulator (SOI) platform using $193\ nm$ lithography technology through a commercial foundry, providing reliable fabrication of feature sizes down to $150\ nm$.

We use a fiber coupled HP 81680A tunable C-band laser providing $2\ mW$ of optical input power. The polarization of the input light is adjusted and maintained with a Thorlabs FPC562 manual polarization controller. Optical coupling into and out of the PIC is achieved via grating couplers designed to be matched to the TE mode of a single-mode fiber angled at $8^\circ$ from the vertical. 

A Thorlabs Nanomax 600 series 6-axis stage is used to precisely align the input and output fibers with their respective grating couplers. The output of the photonic mesh is fiber coupled to an HP 81532A photodetector which has a minimum detectable power of $1\ pW$.

For routing within the photonic integrated circuit, we use standard $500\ nm$ wide strip waveguides with a silicon dioxide cladding. We have designed directional couplers to operate as nominal $50:50$ beam splitters at an operating wavelength of $1550\ nm$. The coupling gap is set at $300\ nm$ to be well within fabrication limitations. The corresponding propagation length of the directional couplers is $40\ \mu{m}$. 

Our phase-shifting elements are implemented as resistive heaters patterned in Titanium Nitride (TiN) above the silicon waveguides, separated by several micrometers of oxide cladding. Thermal isolation trenches have been etched on either side of each heating element to reduce thermal crosstalk between adjacent channels. The global temperature of the PIC is maintained through a TEC directly beneath the host PCB, which is mounted to an aluminum heat sink. 

The electrical inputs of the phase shifters are managed with a National Instruments Digital to Analog Converter (DAC) with 16 bits of precision over $\pm 10\ V$. However, we limit our voltage range to be between $0\ V$ and $5\ V$ to prevent over-driving the heating elements. As a result, our precision is limited to 14 bits. In future work, the driving electronics can be adjusted to recover additional precision; however, modeling of an ideal device via equation \ref{eq:refname1} reveals that 14 bits of resolution may achieve extinction ratios exceeding $80\ dB$. 

Two test devices are initially investigated: one standard MZI and one DMZI. To operate the standard MZI, optical power is directed to the input labeled $E_1$ as given in figure \ref{Schematic}\color{blue}a\color{black}, while the heater power to the phase shifter $\Delta\theta$ is swept and the output power is monitored at detector $D_1$. To locate the phase shifter settings that minimize the optical power at detector $D_1$, a recursive search algorithm was employed.

Initially, the full allowable voltage range from $0\ V$ to $5\ V$ is swept with a coarse resolution. The voltage range and resolution are then successively reduced centering around the optimal state until the full bit depth of the DAC has been reached. A similar sweep is performed for the DMZI test structure after it has been configured according to the algorithm described above in the Methods section. 

\begin{figure}[ht]
\centering
{\includegraphics[width=\linewidth]{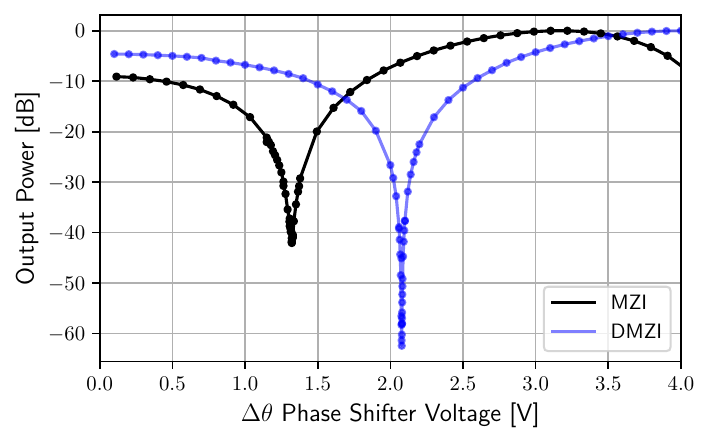}}
\caption{Output power at detector $D_1$ as heater power to phase shifter $\Delta\theta$ in both the standard MZI and the DMZI is swept. The measured extinction ratio for a standard MZI is $42.0\ dB$ corresponding to a split ratio $49.6:50.4$ at each of the beam splitters. The measured extinction ratio for a configured DMZI is $62.4\ dB$ corresponding to a split ratio of $49.96:50.04$ at each of the tunable beam splitters.}
\label{Device_Extinction}
\end{figure} 

The results of this sweep for our test structures are plotted in Figure \ref{Device_Extinction}. The extinction ratio for the standard MZI has been measured as $42.0\ dB$. If it is assumed that the split ratio error is evenly distributed between the two beam splitters of the MZI, the measured extinction ratio corresponds to an error of $\sigma = 7.9*10^{-3}$ or a split ratio of $49.6:50.4$.

The rejection measured for the configured DMZI is $62.4\ dB$, which is in good agreement with prior work on similar devices\cite{Wilkes2016}. This corresponds to an error of $\sigma = 7.6*10^{-4}$ or a split ratio of $49.96:50.04$ representing an order of magnitude reduction to the beam splitter error compared to a standard MZI. While the rejection has been improved significantly by the use of a DMZI compared to a standard MZI, the remaining leakage power is still limiting for high contrast applications. 

\begin{figure*}[ht]
\centering
{\includegraphics[width=\linewidth]{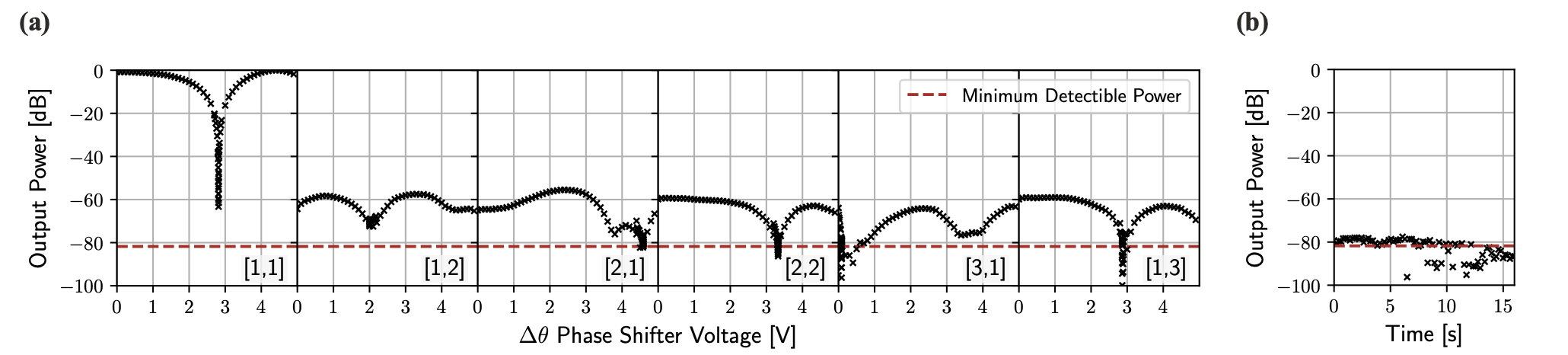}}
\caption{\textbf{(a)}.Relative detected power at the output photodiode $D_1$ of the 3x3 triangular mesh as the internal phase shifter $\Delta\theta$ of each of the 6 devices is tuned sequentially. \textbf{(b)}. After each device has been optimized, the optimal phase shifter settings are held steady while the system reaches a thermal equilibrium. The extinction ratio during this period exceeds $80\ dB$. }
\label{Mesh_Extinction}
\end{figure*}


To achieve better contrast, we use the triangular mesh of DMZIs in figure \ref{Mesh}. Figure \ref{Mesh_Extinction} shows the monitored power at the output, $D_1$, of the DMZI mesh as the internal phase shifter $\Delta\theta$ is tuned at each node. The displayed output power has been normalized by the maximum power measured at the output of the system, $150\ \mu{W}$. The majority of these losses are attributed to the coupling interfaces, with each grating coupler having an estimated insertion loss of $5\ dB$.

As with the standalone MZI and DMZI, the internal phase shifter of each mesh node is first adjusted with a coarse resolution and subsequently refined around the optimal set point. Once a DMZI's optimal setting is achieved, it is held steady while the remaining devices are tuned.

It can be seen in figure \ref{Mesh_Extinction} that optimization of DMZI $[1,1]$ reduces the leakage power to detector $D_1$ by $63.3\ dB$ from the total input power, while the majority of the input power is directed to detector $D_2$. This is in good agreement with testing of the standalone devices and previous works \cite{Wilkes2016}. Optimization of the next DMZI $[1,2]$ along the intended light path further reduces leakage power, deepening the extinction ratio by nearly an order of magnitude to $72.5\ dB$. 


Optimization of DMZI $[2,1]$ further reduces the detected power to less than $1\ pW$ which is the nominal detection limit of our photodiode. This represents a maximum measurable extinction ratio of $81.8\ dB$. The impact of DMZI $[2,1]$ and other devices off of the expected light path is representative of reflections between input and output couplers resulting in the  propagation of light along unexpected paths through the mesh. Continued optimization of the remaining nodes within the mesh results in similar behavior. Apparent rejection beyond $81.8\ dB$ may not be reliable because we are below the nominal detection limit, although the apparent measured values are included in figure \ref{Mesh_Extinction} for completeness.


Following optimization of the entire mesh, the voltage settings applied to each phase shifter in the circuit are kept constant to allow the heating elements and the TEC of the PIC to reach thermal equilibrium. During this period an extinction ratio of $80\ dB$ is maintained.



\textbf{Conclusion.} In this work, we have demonstrated the application of a triangular mesh of double Mach-Zehnder Interferometers (DMZIs) to enhance the extinction ratio in silicon photonic circuits, achieving values exceeding $80\ dB$. By leveraging the additional degrees of freedom provided by the DMZI architecture and employing a systematic tuning algorithm, our approach effectively compensates for fabrication-induced imperfections and non-idealities such as back reflections. This represents a substantial improvement over conventional MZI and standalone DMZI configurations, as evidenced by the measured extinction ratio improvement by two orders of magnitude.

The scalability and adaptability of the triangular DMZI mesh open pathways for its integration into a wide range of high-sensitivity applications, including astronomical interferometry, coronagraphy, and high-contrast optical systems. Moreover, the robustness of the proposed architecture against fabrication errors, combined with its thermal stability, underscores its potential for reliable performance in practical implementations. Future work could explore integration with automated feedback systems for real-time configuration offering a promising direction to simplify the tuning process and enhance operational efficiency.



\begin{backmatter}
\bmsection{Funding} Air Force Office of Scientific Research (FA9550-21-1-0312, FA9550-23-1-0307); Ames Research Center (80NSSC24M0033); Stanford Engineering.

\bmsection{Acknowledgment} We acknowledge many helpful interactions on chip design with Andrea Melloni, Francesco Morichetti, and other researchers at Politecnico di Milano. We also thank Dan Sirbu, Kevin Fogarty, Ruslan Belikov and other researchers at NASA AMES for many insightful discussions.

\bmsection{Disclosures} The authors declare no conflicts of interest.


\bmsection{Data availability} Data underlying the results presented in this paper are not publicly available at this time but may be obtained from the authors upon reasonable request.

\end{backmatter}

\bibliography{sample}

\begin{thebibliography}{10}
\newcommand{\enquote}[1]{``#1''}

\bibitem{Wen2022}
Z.~Wen, Z.~Guan, J.~Dong, \emph{et~al.}, \enquote{A review of sensitivity enhancement in interferometer-based fiber sensors,} {\protect\JournalTitle{Sensors (Basel, Switzerland)}} \textbf{22} (2022).

\bibitem{Huang2021}
T.~Huang, G.~Xu, X.~Tu, \emph{et~al.}, \enquote{Design of highly sensitive interferometric sensors based on subwavelength grating waveguides operating at the dispersion turning point,} {\protect\JournalTitle{J. Opt. Soc. Am. B}} \textbf{38}, 2680--2686 (2021).

\bibitem{Cho2009}
S.-B. Cho and T.-G. Noh, \enquote{Stabilization of a long-armed fiber-optic single-photon interferometer,} {\protect\JournalTitle{Opt. Express}} \textbf{17}, 19027--19032 (2009).

\bibitem{Hati2008}
A.~Hati, C.~Nelson, J.~Taylor, \emph{et~al.}, \enquote{Cancellation of vibration-induced phase noise in optical fibers,} {\protect\JournalTitle{Photonics Technology Letters, IEEE}} \textbf{20}, 1842--1844 (2008).

\bibitem{Bogaerts2020}
W.~Bogaerts, D.~P{\'e}rez, J.~Capmany, \emph{et~al.}, \enquote{Programmable photonic circuits,} {\protect\JournalTitle{Nature}} \textbf{586}, 207--216 (2020).

\bibitem{Capmany2017}
D.~Perez, I.~Gasulla, J.~Fraile-Pelaez, \emph{et~al.}, \enquote{Silicon photonics rectangular universal interferometer,} {\protect\JournalTitle{Laser \& Photonics Reviews}} \textbf{11} (2017).

\bibitem{Bogaerts2018}
W.~Bogaerts and L.~Chrostowski, \enquote{Silicon photonics circuit design: Methods, tools and challenges,} {\protect\JournalTitle{Laser \& Photonics Reviews}} \textbf{12}, 1700237 (2018).

\bibitem{Humphreys2014}
P.~C. Humphreys, B.~J. Metcalf, J.~B. Spring, \emph{et~al.}, \enquote{Strain-optic active control for quantum integrated photonics,} {\protect\JournalTitle{Optics Express}} \textbf{22}, 21719 (2014).

\bibitem{Enright2014}
R.~Enright, S.~Lei, K.~Nolan, \emph{et~al.}, \enquote{A vision for thermally integrated photonics systems,} {\protect\JournalTitle{Bell Labs Technical Journal}} \textbf{19}, 31--45 (2014).

\bibitem{Sun2023}
Z.~Sun, S.~Pai, C.~Valdez, \emph{et~al.}, \enquote{Scalable low-latency optical phase sensor array,} {\protect\JournalTitle{Optica}} \textbf{10}, 1165--1172 (2023).

\bibitem{Pai2023}
S.~Pai, T.~Park, M.~Ball, \emph{et~al.}, \enquote{Experimental evaluation of digitally verifiable photonic computing for blockchain and cryptocurrency,} {\protect\JournalTitle{Optica}} \textbf{10}, 552--560 (2023).

\bibitem{Cheng2021}
J.~Cheng, H.~Zhou, and J.~Dong, \enquote{Photonic matrix computing: From fundamentals to applications,} {\protect\JournalTitle{Nanomaterials}} \textbf{11} (2021).

\bibitem{Annoni2017}
A.~Annoni, E.~Guglielmi, M.~Carminati, \emph{et~al.}, \enquote{Unscrambling light---automatically undoing strong mixing between modes,} {\protect\JournalTitle{Light: Science \& Applications}} \textbf{6}, e17110--e17110 (2017).

\bibitem{Milanizadeh2022}
M.~Milanizadeh, S.~SeyedinNavadeh, F.~Zanetto, \emph{et~al.}, \enquote{Separating arbitrary free-space beams with an integrated photonic processor,} {\protect\JournalTitle{Light: Science \& Applications}} \textbf{11}, 197 (2022).

\bibitem{Vaughan2023}
N.~Desai and {Et al.}, \enquote{Integrated photonic-based coronagraphic systems for future space telescopes,} in \emph{Proc.SPIE,}  vol. 12680 (2023), p. 126801S.

\bibitem{Sirbu2024}
D.~{Sirbu}, R.~{Belikov}, K.~{Fogarty}, \emph{et~al.}, \enquote{{AstroPIC: Architecture options and trades for integrated photonic coronagraphy},} in \emph{American Astronomical Society Meeting Abstracts,}  vol. 243 of \emph{American Astronomical Society Meeting Abstracts} (2024), p. 329.07.

\bibitem{Pai2019}
S.~Pai, B.~Bartlett, O.~Solgaard, and D.~A.~B. Miller, \enquote{Matrix optimization on universal unitary photonic devices,} {\protect\JournalTitle{Phys. Rev. Appl.}} \textbf{11}, 064044 (2019).

\bibitem{Miller2015}
D.~A.~B. Miller, \enquote{Perfect optics with imperfect components,} {\protect\JournalTitle{Optica}} \textbf{2}, 747--750 (2015).

\bibitem{Wilkes2016}
C.~M. Wilkes, X.~Qiang, J.~Wang, \emph{et~al.}, \enquote{60db high-extinction auto-configured mach--zehnder interferometer,} {\protect\JournalTitle{Opt. Lett.}} \textbf{41}, 5318--5321 (2016).

\bibitem{Wang20}
M.~Wang, A.~Ribero, Y.~Xing, and W.~Bogaerts, \enquote{Tolerant, broadband tunable coupler circuit,} {\protect\JournalTitle{Opt. Express}} \textbf{28}, 5555--5566 (2020).

\bibitem{Miller2013}
D.~A.~B. Miller, \enquote{Self-configuring universal linear optical component,} {\protect\JournalTitle{Photon. Res.}} \textbf{1}, 1--15 (2013).

\bibitem{Hamerly22}
R.~Hamerly, S.~Bandyopadhyay, and D.~Englund, \enquote{Asymptotically fault-tolerant programmable photonics,} {\protect\JournalTitle{Nature Communications}} \textbf{13}, 6831 (2022).

\bibitem{Suzuki15}
K.~Suzuki, G.~Cong, K.~Tanizawa, \emph{et~al.}, \enquote{Ultra-high-extinction-ratio 2 {\texttimes} 2 silicon optical switch with variable splitter,} {\protect\JournalTitle{Opt. Express}} \textbf{23}, 9086--9092 (2015).

\end{thebibliography}

\bibliographyfullrefs{sample}

\end{document}